\documentclass[12pt,epsf]{article}
\usepackage{amsmath,amssymb,graphicx,epsfig,latexsym}
\numberwithin{equation}{section}
\newcommand{\be}{\begin{equation}}
\newcommand{\ee}{\end{equation}}
\newcommand{\beqa}{\begin{eqnarray}}
\newcommand{\eeqa}{\end{eqnarray}}
\newcommand{\nn}{\nonumber}

\newcommand{\<}{\langle}

\def\boxit#1{\vbox{\hrule\hbox{\vrule\kern8pt
\vbox{\hbox{\kern8pt}\hbox{\vbox{#1}}\hbox{\kern8pt}}
\kern8pt\vrule}\hrule}}
\def\mathboxit#1{\vbox{\hrule\hbox{\vrule\kern8pt\vbox{\kern8pt
\hbox{$\displaystyle #1$}\kern8pt}\kern8pt\vrule}\hrule}}

\def\IB{\relax\hbox{$\inbar\kern-.3em{\rm B}$}}
\def\IC{\relax\hbox{$\inbar\kern-.3em{\rm C}$}}
\def\ID{\relax\hbox{$\inbar\kern-.3em{\rm D}$}}
\def\IE{\relax\hbox{$\inbar\kern-.3em{\rm E}$}}
\def\IF{\relax\hbox{$\inbar\kern-.3em{\rm F}$}}
\def\IG{\relax\hbox{$\inbar\kern-.3em{\rm G}$}}
\def\IGa{\relax\hbox{${\rm I}\kern-.18em\Gamma$}}
\def\IH{\relax{\rm I\kern-.18em H}}
\def\IK{\relax{\rm I\kern-.18em K}}
\def\IL{\relax{\rm I\kern-.18em L}}
\def\IP{\relax{\rm I\kern-.18em P}}
\def\IR{\relax{\rm I\kern-.18em R}}
\def\IZ{\relax\ifmmode\mathchoice
{\hbox{\cmss Z\kern-.4em Z}}{\hbox{\cmss Z\kern-.4em Z}}
{\lower.9pt\hbox{\cmsss Z\kern-.4em Z}} {\lower1.2pt\hbox{\cmsss
Z\kern-.4em Z}}\else{\cmss Z\kern-.4em Z}\fi}

\def\II{\relax{\rm I\kern-.18em I}}

\def\CP {{\cal P}}

\begin{document}

\setlength{\baselineskip}{7mm}

\vspace{1cm}
\begin{titlepage}

\begin{flushright}

{\tt CERN-PH-TH/2013-193}\\
{\tt NRCPS-HE-36-2013} ~~~~\\

\end{flushright}

\begin{center}
{\Large ~\\{\it Invariant Scalar Product
\\
\vspace{0.3cm}
on
\\
\vspace{0.3cm}
Extended  Poincar\'e Algebra
}

}

\vspace{2cm}

{\sl  George Savvidy\\

\bigskip
\centerline{${}$ \sl Department of Physics, CERN Theory Division CH-1211 Geneva 23, Switzerland}

\bigskip
\centerline{${}$ \sl Demokritos National Research Center, Ag. Paraskevi,  Athens, Greece}
\bigskip

}
\end{center}
\vspace{2cm}

\centerline{{\bf Abstract}}

\vspace{12pt}

\noindent

Two methods can be used
to calculate explicitly the Killing form on a Lie algebra.
The first one is a direct calculation of the traces of the
generators in a matrix representation of the algebra, and the second one is the usage of
the group invariance of the scalar product.
We use both methods in our calculation of  the scalar product on the
extended Poincar\'e algebra $L_G(\CP)$ in order to have a cross check of our results.
The algebra is infinite-dimensional and requires careful treatment of the infinities.
The scalar product on the extended algebra  $L_G(\CP)$ found by both methods
coincides and the important conclusion which follows  is that
Poincar\'e generators are orthogonal to the gauge  generators.

\end{titlepage}
\newpage

\pagestyle{plain}


\section{\it Introduction}

The algebra is defined as follows  \cite{Savvidy:2008zy,Savvidy:2010vb,Antoniadis:2011re}:
\beqa\label{gaugePoincare}
~&&[P^{\mu},~P^{\nu}]=0, \label{extensionofpoincarealgebra}\\
~&&[M^{\mu\nu},~P^{\lambda}] = i(\eta^{\lambda \nu}~P^{\mu}
- \eta^{\lambda \mu }~P^{\nu}) ,\nn\\
~&&[M^{\mu \nu}, ~ M^{\lambda \rho}] = i(\eta^{\mu \rho}~M^{\nu \lambda}
-\eta^{\mu \lambda}~M^{\nu \rho} +
\eta^{\nu \lambda}~M^{\mu \rho}  -
\eta^{\nu \rho}~M^{\mu \lambda} ),\nn\\
\nn\\
\label{levelcommutators}
~&&[P^{\mu},~L_{a}^{\lambda_1 ... \lambda_{s}}]=0,  \\
~&&[M^{\mu \nu}, ~ L_{a}^{\lambda_1 ... \lambda_{s}}] = i(
\eta^{\lambda_1\nu } L_{a}^{\mu \lambda_2... \lambda_{s}}
-\eta^{\lambda_1\mu} L_{a}^{\nu\lambda_2... \lambda_{s}}
+...+
\eta^{\lambda_s\nu } L_{a}^{\lambda_1... \lambda_{s-1}\mu } -
\eta^{\lambda_s\mu } L_{a}^{\lambda_1... \lambda_{s-1}\nu } ),\nonumber\\
\nn\\
\label{alakac}
~&&[L_{a}^{ \lambda_1 ... \lambda_{i}}, L_{b}^{\lambda_{i+1} ... \lambda_{s} }]=if_{ab}^{c}
L_{c}^{\lambda_1 ... \lambda_{s} }       ~~~(\mu,\nu,\rho,\lambda=0,1,2,3; ~~~~~s=0,1,2,... ),
\eeqa
where the flat space-time metric is
$
\eta^{\mu \nu}= diag(+1,-1,-1,-1).
$
One can check that all Jacoby identities are satisfied and we have an example of a
fully consistent algebra.  The algebra $L_G(\CP)$ incorporates
the Poincar\'e algebra $L_{\CP}$ and an internal algebra $L_G $ in a nontrivial way, which is different from
the direct product. The generators $L_{a}^{\lambda_1 ... \lambda_{s}}$
have a nonzero commutation relation with $  M^{\mu\nu} $,
which means that the generators of this new symmetry have nontrivial Lorentz
transformation and carry a spin different from zero.
The generators $L_{a}^{\lambda_1 ... \lambda_{s}}$ in (\ref{alakac})   commute to themselves
forming an infinite series of commutators of  current  subalgebra
which cannot be truncated, so that the index $s$ runs from zero to infinity. We have here an example
of an infinitely-dimensional current subalgebra   \cite{Faddeev:1984jp,Goddard:1986bp}.

The algebra is invariant with respect to the following gauge transformations:
\beqa\label{isomorfism}
& L_{a}^{\lambda_1 ... \lambda_{s}} \rightarrow L_{a}^{\lambda_1 ... \lambda_{s}}
+ \sum_{1} P^{\lambda_1}L_{a}^{\lambda_2 ... \lambda_{s}}+
\sum_{2} P^{\lambda_1} P^{\lambda_2} L_{a}^{\lambda_3 ... \lambda_{s}} +...+
P^{\lambda_1}... P^{\lambda_s} L_{a}\\
& M^{\mu\nu} \rightarrow M^{\mu\nu},~~~~
P^{\lambda} \rightarrow P^{\lambda},\nn
\eeqa
where the sums  $\sum_{1},\sum_{2},... $ are over all inequivalent index permutations.
The algebra $L_G(\CP)$ has representation in terms of differential operators of the form:
\beqa\label{represofextenpoincarealgebra}
~&& P^{\mu} = k^{\mu} ,\nn\\
~&& M^{\mu\nu} = i(k^{\mu}~ {\partial\over \partial k_{\nu}}
- k^{\nu }~ {\partial \over \partial k_{\mu}}) + i(e^{\mu}~ {\partial\over \partial e_{\nu}}
- e^{\nu }~ {\partial \over \partial e_{\mu}}),\nn\\
~&& L_{a}^{\lambda_1 ... \lambda_{s}} =e^{\lambda_1}...e^{\lambda_s} \otimes L_a,
\eeqa
where $e^{\lambda} \in M^4$ is a translationally invariant space-time vector.

It is worthwhile to compare the above extension of the Poincar\'e algebra with
the one which has been
considered long ago by Ogievetski, Ivanov  and others in a series of articles
\cite{Borisov:1974bn,Ivanov:1976zq,Ivanov:1976pg,Ivanov:1979ny}. These authors were
developing the idea that gauge bosons such as the  photon, the Yang-Mills quanta
and the graviton are Goldstone particles and  therefore the
spontaneously broken symmetry is more profound and general concept than
the gauge symmetry \cite{Ivanov:1976zq}. In their approach the gauge transformations were considered
as constant parameter transformations of a group which has an infinite number of generators
$Q_a^{\mu_1 ... \mu_{n}}= x^{\mu_1}...x^{\mu_n} L_a$ together with the Poincar\'e
generators
\beqa
&P^{\mu}=i  {\partial\over \partial x_{\mu}}~~,~~~~
M^{\mu\nu}= i(x^{\mu}{\partial\over \partial x_{\nu}}  -x^{\nu}{\partial\over \partial x_{\mu}})\nn\\
&Q_a^{\mu_1 ... \mu_{n}}= x^{\mu_1}...x^{\mu_n} L_a~.\nn
\eeqa
The generators $Q_a^{\mu_1 ... \mu_{n}}$ are not
translationally invariant because $[P^{\mu}, Q_a^{\mu_1 ... \mu_{n}}] \neq 0$,
therefore they are essentially different from the generators $L_{a}^{\lambda_1 ... \lambda_{s}}$
with their vanishing  commutators
$[P^{\mu}, L_{a}^{\lambda_1 ... \lambda_{s}}] = 0$
in (\ref{levelcommutators}).  Still there is  a similarity
between the generators $L_{a}^{\lambda_1 ... \lambda_{s}}$ and $Q_a^{\mu_1 ... \mu_{n}}$.

An example of the infinite-dimensional algebra which contains Lorentz subalgebra and
the high-rank multispinor generators $Q^{\alpha_1...\alpha_n \dot{\beta}_1...\dot{\beta}_m}$ were
considered in the article of Vasiliev \cite{Vasiliev:1995dn}. It is a generalization of the
$L_{sp(4)}$ algebra to the infinite-dimensional associative algebra $L_{hs(4)}$ spanned by
high order polynomials constructed from mutually conjugated spinor oscillators. In this
algebra there are no
internal charges $L_a$ associated with the generators $Q^{\alpha_1...\alpha_n \dot{\beta}_1...\dot{\beta}_m}$.
The details of the
calculations concerning the traces of the generators on $L_{hs(4)}$ can be found in
\cite{Vasiliev:1995dn,Doroud:2011xs}.

For the purposes of constructing
field-theoretical models based on a larger group of symmetry one should
define the notion of a trace for the fields taking
values on the extended algebra and in particular on   $L_G(\CP)$.
In the next section we shall recollect the useful formulae defining the structure of the invariant
scalar product both for the compact and noncompact finite-dimensional Lie algebras and in
particular the invariant scalar product of the Poincar\'e generators.  Two methods can be used
to calculate explicitly the Killing form. The first one is a direct calculation of the traces of the
generators in a matrix representation of the algebra, and the second one is the usage of
the group invariance of the scalar product.
We shall use both methods in our calculation of  the scalar product on the
extended Poincar\'e algebra $L_G(\CP)$ in order to have a cross check of our final result. The reason
is that the algebra is infinite-dimensional and requires careful treatment of the infinities.
In the third section we shall use the explicit matrix representation of the $L_G(\CP)$ algebra
generators to calculate the traces and in the fourth section we shall use group invariance of the scalar product.
The results for the scalar product on the extended algebra  $L_G(\CP)$ found by both methods
coincide and are presented in the following table:
\beqa
L_G:~~~~~~&&~~~~ \<L_{a}; L_{b} \rangle  =\delta_{ab}, \label{finale0}\\
\nn\\
L_{\CP}:~~~~~&& ~~~\<P^{\mu} ; P^{\nu }  \rangle ~=0\nn\\
&&~~~\<M_{\mu\nu} ; P_{\lambda  }  \rangle ~  =0\label{poincare0}\\
&& ~~~\<M^{\mu\nu} ; M^{\lambda \rho }  \rangle =\eta^{\mu\lambda } \eta^{\nu\rho}
-\eta^{\mu\rho} \eta^{\nu \lambda }\nn\\
\nn\\
L_G(\CP):~~~~~&&~~~~~~\<P^{\mu};L_{a}^{ \lambda_1 ... \lambda_{s}}\rangle  =0,\nn\\
&&~~~~~~\<M^{\mu\nu};L_{a}^{ \lambda_1 ... \lambda_{s}}\rangle =0, \label{poincarecurrent0}\\
\nn\\
&&~~~~~~\<L_{a}; L^{\lambda_1}_{b} \rangle  =0,\nn\\
&&~~~~~~\<L^{\lambda_1}_{a}; L^{\lambda_2}_{b} \rangle  = \delta_{ab}~ \eta^{\lambda_1 \lambda_2} ,\nn\\
&&~~~~~~\<L_{a}; L^{\lambda_1\lambda_2}_{b} \rangle  = \delta_{ab}~ \eta^{\lambda_1 \lambda_2} ,\nn\\
&&~~~~~~\<L^{\lambda_1}_{a}; L^{\lambda_2 \lambda_3}_{b} \rangle  =0,\label{currentscalar0}\\
&&~~~~~~~~~~.....................\nn\\
&&~~~~~~\<L^{\lambda_1...\lambda_n}_{a}; L^{\lambda_{n+1}....\lambda_{2s+1}}_{b} \rangle  = 0,~~~~~~~~~s=0,1,2,3,...\nn\\
&&~~~~~~\<L^{\lambda_1...\lambda_n}_{a}; L^{\lambda_{n+1}....\lambda_{2s}}_{b} \rangle  =
\delta_{ab}~ s!~(\eta^{\lambda_1 \lambda_2}  \eta^{\lambda_3 \lambda_4}...
\eta^{\lambda_{2s-1} \lambda_{2s}} +\textrm{perm}).\nn
\eeqa
The Killing forms on the internal $L_G$  and on the Poincar\'e  $L_{\CP}$
 subalgebras are well known (\ref{finale0}), (\ref{poincare0}).
The important conclusion which follows from the above result is that
Poincar\'e generators $P^{\mu}, M^{\mu\nu}$ are orthogonal to the  gauge  generators
$L_{a}^{ \lambda_1 ... \lambda_{s}}$ (\ref{poincarecurrent0}). The last formulas
(\ref{currentscalar0}) represent the Killing form on the current algebra (\ref{alakac}).

\section{\it Invariant Killing Forms  }

Our intension is to define  the invariant scalar product on the extended
algebra $L_G(\CP)$. Let us recollect the structure of the invariant scalar product
in the cases of finite-dimensional Lie algebras $L_G$ \cite{Barut:1986}. The scalar product of arbitrary two
elements $X=X^a L_a$ and $Y=Y^a L_a$ of the algebra  is defined  as a trace of the
generators in the adjoint representation $(L_a)^c_d =if_{ad}^{c} $~:
\be\label{adjoint}
\<X;Y\rangle  = tr(ad~ X ~ad~ Y) =X^a (L_a)^c_d ~~Y^b (L_b)^d_c =X^a  if_{ad}^{c} ~Y^b i  f^d_{bc}=g_{ab}~X^a Y^b~,
\ee
where the associated Cartan metric $g_{ab}$ is
\be\label{cartanmetric}
g_{ab} =if_{ad}^{c} ~i  f^d_{bc}~.
\ee
If X and Y are the single generators we shall have
\be
\<L_a;L_b\rangle = g_{ab}.
\ee
The scalar product, if defined as a trace of the generators, depends only on the scales
set by choice of a given representation and not on a particular representation used,
therefore it is convenient to  take the generators in the adjoint representation and express the scalar
product in terms of the structure constants, as it is in (\ref{adjoint}) and (\ref{cartanmetric}).

The last two expressions can be used to find the explicit form
of the scalar product for a specific algebra.
So defined scalar product is symmetric and bilinear:
$$
\<X;Y\rangle  =\<Y;X\rangle ,~~\<\alpha X + \beta Y;Z\rangle = \alpha \<X;Z\rangle  +   \beta \<Y;Z\rangle ,
$$
where $X=X^a L_a,~ Y=Y^a L_a,~Z=Z^a L_a$
and is invariant under the action of the group
$$
\<g X g^{-1}; g Y g^{-1}\rangle = \<X;Y \rangle ~~~~~~~~~~~~~~g \in G~~.
$$
For infinitesimal group elements it is equivalent to
the expression \cite{Goddard:1986bp,Barut:1986}:
\be\label{symmetic}
\<[X,Y]; Z]\rangle  + \<Y;[X, Z]\rangle =0 ~~~~~~X,Y,Z \in L_G~,
\ee
which can also  be used to define the invariant scalar product on the algebra. In the subsequent
sections we shall use both methods to find scalar product on the extended algebra
 $L_G(\CP)$. Because the representations of this algebra are infinite-dimensional,
it is important to have an independent check by both methods, that is, by direct computation
of traces (\ref{adjoint}) and by using the group invariance of the scalar product (\ref{symmetic}).
\vspace{0,1cm}

As an example let us consider the general linear algebra $L_{gl(n)}$   which is
defined by the following commutation relation \cite{Barut:1986}:
\be
[L_{ij}, L_{kl}]= \delta_{jk}~L_{il}-\delta_{il}~L_{kj},
\ee
i,j,k,...=1,2,...,n, with its structure constants
\be
if^{sm}_{ij,kl}=\delta^{s}_{i}\delta_{jk}\delta^{m}_{l}-
\delta^{s}_{k}\delta_{il}\delta^{m}_{j},
\ee
so that for the Cartan metric (\ref{cartanmetric}) one can get
\be
\<L_{ij};L_{pr}\rangle =g_{ij, pr}= if^{sm}_{ij,kl}~ if^{kl}_{pr,sm} =
2n \delta_{ir}\delta_{jp}-2 \delta_{ij}\delta_{pr},
\ee
or, equivalently,
\be
\<X;Y\rangle  = \<X^{ij} L_{ij}; Y^{pr} L_{pr}\rangle = 2n~ tr(XY)-2 tr(X) tr(Y).
\ee
\vspace{0,1cm}

For the $L_{so(p,q)}$  algebra with its commutation relation
\be
[M_{AB}, M_{CD}]= \eta_{AD}~M_{BC}- \eta_{AC}~M_{BD}+
\eta_{BC}~M_{AD}- \eta_{BD}~M_{AC},
\ee
where
$
\eta_{AB}= diag(+1,...,+1,-1,...,-1)
$
is a diagonal matrix with p minuses and q pluses, we have
$$
if^{KL}_{AB,CD}= \delta^K_B \eta_{AD} \delta^L_C - \delta^K_B \eta_{AC} \delta^L_D+
\delta^K_A \eta_{BC} \delta^L_D-\delta^K_A \eta_{BD} \delta^L_C~,
$$
and using (\ref{cartanmetric}) one can get the following expression for the Cartan
metric \cite{Barut:1986}:
\be\label{invscalso}
\<M_{AB};M_{CD}\rangle =g_{AB,CD}~=~(2p+2q-4)(\eta_{AD}\eta_{BC}- \eta_{AC}\eta_{BD}),
\ee
or, equivalently,
\be
\<X;Y\rangle =\<X^{AB}M_{AB};Y^{CD}M_{CD}\rangle ~=~(p+q-2)~tr(X-\tilde{X}  )\cdot(Y-\tilde{Y}).
\ee
\vspace{0,1cm}

Because the Poincar\'e algebra $L_{\CP}$ can be considered as a contraction of the
de Sitter algebra $L_{so(3,2)}$ with $\eta_{AB}= diag(+---+)$, A,B=(0,1,2,3,4) by taking
$M_{4\mu}= R P_{\mu}$ and $R \rightarrow \infty$, where $\mu,\nu,... = (0,1,2,3)$,
we can calculate the scalar product on $L_{\CP}$ using invariant scalar product
(\ref{invscalso}):
\beqa
\<P_{\mu};P_{\nu}\rangle = {1\over R^2} \<M_{4\mu};M_{4\nu}\rangle = -{6\over R^2} \eta_{44}\eta_{\mu\nu}\rightarrow 0\nn\\
\<M_{\mu\nu};P_{\lambda}\rangle = {1\over R} \<M_{\mu\nu};M_{4\lambda}\rangle = {6\over R}
(\eta_{\mu\lambda}\eta_{\nu 4}-\eta_{\nu\lambda}\eta_{\mu 4}) = 0\nn\\
\<M_{\mu\nu};M_{\lambda\rho}\rangle =
6 (\eta_{\mu\lambda}\eta_{\nu \rho}-\eta_{\nu\lambda}\eta_{\mu \rho}) = 0.\nn
\eeqa
Thus the scalar product of the Poincar\'e generators is defined as follows:
\beqa\label{killingform0}
&& ~~~\<P_{\mu} ; P_{\nu }  \rangle ~=0\nn\\
L_{\CP}:~~~~~&&~~~\<M_{\mu\nu} ; P_{\lambda  }  \rangle ~  =0\\
&& ~~~\<M_{\mu\nu} ; M_{\lambda \rho }  \rangle =\eta_{\mu\lambda } \eta_{\nu\rho}
-\eta_{\mu\rho} \eta_{\nu \lambda }\nn
\eeqa

If the Cartan metric $g_{ab}$ of the semi-simple algebra $L_G$ is positive definite,
then the corresponding algebra is compact and one can choose the basis of the
generators $\{ L_a \}$ of $L_G$ so that $g_{ab}= \delta_{ab}$ and define
the covariant rank-3 tensor as
$$
f_{abc}= f^{d}_{ab} ~g_{dc}= f^{c}_{ab}.
$$
It is antisymmetric over (a,b) and, as one can see from the relation
$$
f_{abc}= f^{d}_{ab} ~g_{dc} = f^{d}_{ab} ~ f^n_{d m}~ f^m_{cn}=
f^{n}_{ad} ~f^d_{bm}~ f^m_{cn}+f^{n}_{bd} ~f^d_{ma}~ f^m_{cn},
$$
it is symmetric over cyclic permutations of (a,b,c)\footnote{One should use the Jacobi identity
$f^{d}_{ab} ~ f^n_{d m}= - f^n_{md} ~ f^{d}_{ab}~ = ~f^{n}_{ad} ~f^d_{bm}~+~f^{n}_{bd} ~f^d_{ma}$.},
and thus it is totally antisymmetric.
In summary, for a compact semi-simple Lie algebra $L_G$
one can choose a basis of the
generators $\{ L_a \}$ such that the generators are orthogonal \cite{Barut:1986}
\be\label{simplegroup}
\<L_a;L_b\rangle  = \delta_{ab}
\ee
and the structure constants $f_{abc}$ are totally antisymmetric.

\section{\it Killing Form on $L_G(\CP)$ Algebra}

Now let us define  the invariant scalar product on the extended
algebra $L_G(\CP)$. As we mentioned above, there are two ways by which we
can define a scalar product.
The first one is a direct calculation of the traces of the
generators in the matrix representations of the algebra as in (\ref{adjoint}),
and the second one is the usage of
the group invariance of the scalar product (\ref{symmetic}).
We shall proceed with the explicit matrix
representation of the $L_G(\CP)$ generators. These representations have been constructed in
\cite{Savvidy:2010vb,Antoniadis:2011re}.  It has the representation  of the following
form:
\beqa\label{represofextenpoincarealgebra}
~&& P^{\mu} = k^{\mu} ,\nn\\
~&& M^{\mu\nu} = i(k^{\mu}~ {\partial\over \partial k_{\nu}}
- k^{\nu }~ {\partial \over \partial k_{\mu}}) + i(\xi^{\mu}~ {\partial\over \partial \xi_{\nu}}
- \xi^{\nu }~ {\partial \over \partial \xi_{\mu}})  \\
~&& L_{a}^{\lambda_1 ... \lambda_{s}} =\xi^{\lambda_1}...\xi^{\lambda_s} \otimes L_a ,\nn
\eeqa
where the vector space is parameterized
by momentum coordinates $k^{\mu}$ and translationally invariant vector variables $\xi^{\mu}$:
\be\label{vectorspace}
\Psi(k^{\mu}, \xi^{\nu} )~.
\ee
The irreducible representations can be obtained from (\ref{represofextenpoincarealgebra})
by   imposing  invariant constraints on the
vector space of functions  (\ref{vectorspace})
of the following form \cite{wigner,yukawa1,fierz,wigner1}:
\be\label{constraint}
k^2=0,~~~k^{\mu} \xi_{\mu}=0,~~~\xi^2=-1~.
\ee
These equations have a unique solution
\be\label{solution}
\xi^{\mu}= \xi k^{\mu} + e^{\mu}_{1}\cos\varphi +e^{\mu}_{2}\sin\varphi,
\ee
where $e^{\mu}_{1}=(0,1,0,0),~ e^{\mu}_{2}=(0,0,1,0)$ when $k^{\mu}=\omega(1,0,0,1)$. The
invariant subspace of functions (\ref{represofextenpoincarealgebra}) now reduces to
the form
\be\label{independentvariables}
\Psi(k^{\mu}, \xi^{\nu} )~\delta(k^2)~\delta(k\cdot\xi)~\delta(\xi^2 +1)
= \Phi(k^{\mu}, \varphi, \xi,  ),
\ee
where $\xi$ and $\varphi$ remain as independent variables
on the cylinder $ \varphi \in S^1, \xi \in R^1 $.
The generators of the little group L
\cite{wigner,wigner1, Brink:2002zx}, which leave the fixed momentum $k^{\mu}=k(1,0,0,1)$ invariant,
form the $E(2)$ algebra:
\beqa
[h,\pi^{'}]=
i\pi^{''},~~~[h,\pi^{''}]= -i\pi^{'},~~~[\pi^{'},\pi^{''}]=0~,\nn
\eeqa
where
$
h= M_{12},~~~\pi^{'}= M_{10} + M_{13} ,~~~\pi^{''} = M_{20} + M_{23}.\nn
$
Notice that the transformations which are generated by $L_{a}^{ \lambda_1 ... \lambda_{s}}$
also leave the manifold of states with fixed
momentum  invariant, so that we have to add them to the little algebra L
\cite{Savvidy:2010vb,Antoniadis:2011re}:
$$
h,~~~\pi^{'} ,~~~\pi^{''}, ~~~L_{a}^{ \lambda_1 ... \lambda_{s}}.
$$
The representation of the little  algebra L in terms of differential operators is of the form
\beqa\label{little}
&h= -i{\partial \over \partial\varphi}  , \nn\\
&\pi^{'}= \rho \cos\varphi ,~~~~
\pi^{''}=\rho \sin\varphi  ,~~~~
\rho = -{i\over \omega} {\partial \over \partial \xi},
\eeqa
and taking into account (\ref{solution}) the $L_{a}^{ \lambda_1 ... \lambda_{s}}$ generators take the form
\be\label{trasversalgenera}
L_{a}^{\bot~ \mu_1 ... \mu_{s}}= \prod^{s}_{i=1} ( \xi k^{\mu_i} + e^{\mu_i}_{1}\cos\varphi
+e^{\mu_i}_{2}\sin\varphi)\oplus L_a.
\ee
This is a purely transversal representation  in the sense that
\be
k_{\lambda_1}L_{a}^{\bot \lambda_1 ... \lambda_{s}}=0,~~~~s=1,2,...
\ee
Below we shall essentially use the operator representation  (\ref{little}) and (\ref{trasversalgenera})
to calculate matrix elements and traces of the operator products. It is also important to know
the helicity content of the gauge operators $L_{a}^{ \lambda_1 ... \lambda_{s}}$.
The Poincar\'e generators $\pi^{\pm} = \pi^{'} \pm \pi^{''}$ carry
helicities $h=(1,-1)$. The  $L^{\pm}_a= L^{1}_{a} \pm i L^{2}_{a}$ carry helicities
$h=(1,-1)$, as seen from
$
[h,~L^{\pm}_a] = \pm L^{\pm}_a.
$
The rank-2 generators $L^{++}_a, L^{+-}_a,L^{--}_a$
$$
L^{++}_a = L^{11}_a +2i L^{12}_a - L^{22}_a,~~~~~L^{+-}_a = L^{11}_a + L^{22}_a,~~~~~
L^{--}_a = L^{11}_a -2i L^{12}_a -L^{22}_a,
$$
carry helicities $h=(2,0,-2)$ because  $[h, ~L^{\pm\pm}_a] = \pm 2  L^{\pm\pm}_a,~~[h, ~L^{+-}_a]=0 $.
In general the rank-s
$(L^{+\cdot\cdot\cdot+}_{a},...,L^{-\cdot\cdot\cdot-}_{a})$ generators  carry helicities in the following range:
\be\label{trasversalgenera1}
h=(s,s-2,......, -s+2, -s),
\ee
in total $s+1$ states. This result proves that the algebra $L_G(\CP)$ has representations
which describe propagation of the high helicity charged states.
\\

Having in hand the explicit representation of the $L_G(\CP)$ generators we can find their
matrix elements  and calculate the corresponding traces of the generators. As a basis of
functions on a cylinder $ \varphi \in S^1, \xi \in R^1 $  we shall take (see Appendix A for details)
\beqa
\vert n,m\rangle = {1\over \sqrt{2\pi}} e^{i n \varphi} ~~ {\exp{(-\xi^2/2)} \over \sqrt{2^m m! \sqrt{\pi}}}H_m(\xi)
={1\over \sqrt{2\pi}} e^{i n \varphi} ~~\psi_m(\xi),~\nn
\eeqa
where  $H_m(\xi)$ are Hermite polynomials and the trace should be defined as
\beqa\label{inftytrace}
\<A;B\rangle ~&=&~\sum_{n,m} \int^{\pi}_{-\pi}\int^{\infty}_{-\infty} d\varphi d \xi \<n,m\vert A B \vert n,m\rangle \nn\\
&=&\sum_{n,m}\sum_{k,l}\<n,m\vert A \vert k,l\rangle  \<k,l \vert B \vert n,m\rangle .
\eeqa
The traces of the  Poincar\'e generators (\ref{little}) can be easily computed:
\beqa
\<h;\pi^{'}\rangle =\sum_{n,m}\sum_{k,l}\<n,m\vert h \vert k,l\rangle  \<k,l \vert \pi^{'} \vert n,m\rangle =\nn\\
=\sum_{n,m}\sum_{k,l} k \delta_{n,k} \delta_{m,l} ~{1\over 2}(\delta_{k,n+1} +\delta_{k,n-1})
 (\sqrt{2 m}~\delta_{l, m-1} -\sqrt{2 (m+1)}\delta_{l, m+1})=0, \nn
\eeqa
and in a similar way $\<h;\pi^{'}\rangle =\<\pi^{'};\pi^{''}\rangle =0$.
This explicit calculation confirms the previous result (\ref{killingform0}),
which we had for the Poincar\'e generators. Indeed, the Little algebra generators
are defined as $h= M_{12},~~~\pi^{'}= M_{10} + M_{13} ,~~~\pi^{''} = M_{20} + M_{23} $
and if one takes into account that the scalar product between $M_{\mu\nu}$ generators is given
by (\ref{killingform0}) we shall see that $h, \pi^{'}$ and $\pi^{''}$ are indeed orthogonal.

Now we are prepared to calculate the traces between Poincar\'e generators and the
gauge  generators  $L_{a}^{ \lambda_1 ... \lambda_{s}}$. We have
\beqa
&\<h;L_{a}\rangle =\<n,m\vert h \vert r,l\rangle  \<r,l \vert L_a \vert n,m\rangle =\nn\\
&=k \delta_{n,r} \delta_{m,l} ~\delta_{r,n}\delta_{m,l}\<1;L_a\rangle =0,\nn
\eeqa
that is they are orthogonal. For the vector generator $L^{\mu}_{a}$ we shall get
\beqa
 \<h;L^{\mu}_{a}\rangle =\<n,m\vert h \vert r,l\rangle  \<r,l \vert
\xi k^{\mu} + e^{\mu}_{1}\cos\varphi
+e^{\mu}_{2}\sin\varphi \vert n,m\rangle \<1;L_a\rangle =\nn\\
 =r \delta_{n,r} \delta_{m,l}~  \{k^{\mu}(\sqrt{{m\over 2}} ~\delta_{l, m-1} +\sqrt{{(m+1)\over 2}} \delta_{l, m+1})
\delta_{r,n} +
 (e^{\mu}_{+} \delta_{r,n+1} + e^{\mu}_{-} \delta_{r,n-1})\delta_{l,m}
\}\<1;L_a\rangle =0,\nn
\eeqa
where $2 e^{\mu}_{\pm} = e^{\mu}_{1} \mp i e^{\mu}_{2}$
and in general one can get convinced that generators $h, \pi^{'},\pi^{''}$ and gauge generators
$L_{a}^{ \lambda_1 ... \lambda_{s}}$ are orthogonal to each other:
\beqa
&&~~~~~\<h;L_{a}^{ \lambda_1 ... \lambda_{s}}\rangle  =0,\nn\\
L:~~~~~&&~~~~\<\pi^{'};L_{a}^{ \lambda_1 ... \lambda_{s}}\rangle =0,\\
&&~~~~~\<\pi^{''};L_{a}^{ \lambda_1 ... \lambda_{s}}\rangle  =0.\nn
\eeqa
This statement can be extended to all Lorentz generators since a state of the Hilbert space
with fixed momentum $k^{\mu}$ will transform to the state with momentum $k^{'} = \Lambda k $
if one applies the group operator  $U_\Lambda$
corresponding to the Lorentz transformation $\Lambda_{\mu\nu}$. Therefore we have
\beqa
L_G(\CP):~~~~~&&~~~~\<M^{\mu\nu};L_{a}^{ \lambda_1 ... \lambda_{s}}\rangle =0.
\eeqa
Finally we have to calculate the scalar product between gauge  generators $L_{a}^{ \lambda_1 ... \lambda_{s}}$.
For the compact Lie algebra $L_G$ we have (\ref{simplegroup})
\beqa
L_G:~~~~~~~~~~~~ \<L_{a}; L_{b} \rangle  =\delta_{ab},\nn
\eeqa
and then with the first level generator $L^{\mu}_{b}$
\beqa
&\<L_{a}; L^{\mu}_{b} \rangle  = \<n,m \vert r,l\rangle  \<r,l \vert
\xi k^{\mu} + e^{\mu}_{1}\cos\varphi
+e^{\mu}_{2}\sin\varphi \vert n,m\rangle \<L_a;L_b\rangle =\nn\\
&= \delta_{n,r} \delta_{m,l}~  \{k^{\mu}(\sqrt{{m\over 2}} ~\delta_{l, m-1} +\sqrt{{(m+1)\over 2}} \delta_{l, m+1})
\delta_{r,n}
 + (e^{\mu}_{+} \delta_{r,n+1} + e^{\mu}_{-} \delta_{r,n-1})\delta_{l,m}
\}\delta_{ab} =0.\nn
\eeqa
The scalar product between first and third level generators also nullifies, as one can see from
the following calculation:
\beqa
&\<L^{\mu}_{a}; L^{\nu \rho}_{b} \rangle  = \<n,m \vert
\xi k^{\mu} + e^{\mu}_{1}\cos\varphi
+e^{\mu}_{2}\sin\varphi \vert r,l\rangle  \nn \\
&\<r,l \vert
(\xi k^{\nu} + e^{\nu}_{1}\cos\varphi
+e^{\nu}_{2}\sin\varphi) (\xi k^{\rho} + e^{\rho}_{1}\cos\varphi
+e^{\rho}_{2}\sin\varphi) \vert n,m\rangle  \<L_a;L_b\rangle =\nn\\
&=  \{k^{\mu}(\sqrt{{l\over 2}} ~\delta_{m, l-1} +\sqrt{{(l+1)\over 2}} \delta_{m, l+1})
\delta_{n,r}
 +( e^{\mu}_{+} \delta_{n,r+1} + e^{\mu}_{-} \delta_{n,r-1})
\delta_{m,l}\} \times \nn\\
&\{k^{\nu} k^{\rho}(\sqrt{{m(m-1)\over 4}} ~\delta_{l, m-2}+{2m +1\over 2}  ~\delta_{l, m }
+\sqrt{{(m+1)(m+2)\over 4}} \delta_{l, m+2})
\delta_{r,n} +  \\
& +k^{\nu}(\sqrt{{m\over 2}} ~\delta_{l, m-1} +\sqrt{{(m+1)\over 2}} \delta_{l, m+1})
( e^{\rho}_{+} \delta_{r,n+1} + e^{\rho}_{-} \delta_{r,n-1}) +\nn\\
&+( e^{\nu}_{+} \delta_{r,n+1} + e^{\nu}_{-} \delta_{r,n-1})
k^{\rho}(\sqrt{{m\over 2}} ~\delta_{l, m-1} +\sqrt{{(m+1)\over 2}} \delta_{l, m+1})+\nn\\
&+ \delta_{l,m} (e^{\nu}_{+}e^{\rho}_{+} \delta_{r,n+2} + (e^{\nu}_{+}e^{\rho}_{-}+ e^{\nu}_{-} e^{\rho}_{+}) \delta_{r,n}
+e^{\nu}_{-}e^{\rho}_{-} \delta_{r,n-2}) \}\delta_{ab} = 0.\nn
\eeqa
Generally when the total number of Lorentz indices
in the scalar product is odd
one finds that the scalar product vanishes:
$\<L^{\lambda_1...\lambda_n}_{a}; L^{\lambda_{n+1}....\lambda_{2s+1}}_{b} \rangle  = 0$.
The scalar product is nonzero when the number of indices is even\footnote{The sum in the trace
can be regularized
by defining the traces of the infinite-dimensional matrices (\ref{inftytrace}) by the exponential
weight factor $\exp{(-n^2 -m^2)}$.}:
\beqa
&\<L^{\mu}_{a}; L^{\nu}_{b} \rangle  = \<n,m \vert
\xi k^{\mu} + e^{\mu}_{1}\cos\varphi
+e^{\mu}_{2}\sin\varphi \vert r,l\rangle  \nn \\
&\<r,l \vert
\xi k^{\nu} + e^{\nu}_{1}\cos\varphi
+e^{\nu}_{2}\sin\varphi \vert n,m\rangle  \<L_a;L_b\rangle =\nn\\
&=  \{k^{\mu}(\sqrt{{l\over 2}} ~\delta_{m, l-1} +\sqrt{{(l+1)\over 2}} \delta_{m, l+1})
\delta_{n,r}
 +( e^{\mu}_{+} \delta_{n,r+1} + e^{\mu}_{-} \delta_{n,r-1})
\delta_{m,l}\}\nn\\
&\{k^{\nu}(\sqrt{{m\over 2}} ~\delta_{l, m-1} +\sqrt{{(m+1)\over 2}} \delta_{l, m+1})
\delta_{r,n} +
 (e^{\nu}_{+} \delta_{r,n+1} + e^{\nu}_{-} \delta_{r,n-1})
\delta_{l,m}\}\delta_{ab} = \nn\\
&=\delta_{ab} \sum_{n,m} \{k^{\mu}k^{\nu}  (m+1/2) + e^{\mu}_{+}e^{\nu}_{-}+e^{\mu}_{-}e^{\nu}_{+}\}.
\eeqa
We should also average over all orientations of the momentum $k^{\mu}$. Its rotation is generated by
the application of the group operator $U_\Lambda $
corresponding to the Lorentz rotation with the group parameters $\Lambda_{\mu\nu}$. This  average
in nonzero and proportional to the  $\eta^{\mu\nu}$, thus
we find that
\be\label{currentgen}
\<L^{\mu}_{a}; L^{\nu}_{b} \rangle  = \delta_{ab} \eta^{\mu\nu}.
\ee
In a similar way
\be\label{currentgen1}
\<L_{a}; L^{\mu\nu}_{b} \rangle  = \delta_{ab} \eta^{\mu\nu}.
\ee
We can summarize now the structure of the scalar product of the extended algebra  $L_G(\CP)$ in the following
table:
\beqa\label{finale}
L_G:~~~~~~&&~~~~~~~ \<L_{a}; L_{b} \rangle  =\delta_{ab}, \nn\\
\nn\\
L_{\CP}:~~~~~&& ~~~\<P^{\mu} ; P^{\nu }  \rangle ~=0\nn\\
&&~~~\<M_{\mu\nu} ; P_{\lambda  }  \rangle ~  =0\label{poincare}\\
&& ~~~\<M^{\mu\nu} ; M^{\lambda \rho }  \rangle =\eta^{\mu\lambda } \eta^{\nu\rho}
-\eta^{\mu\rho} \eta^{\nu \lambda }\nn\\
\nn\\
L_G(\CP):~~~~~&&~~~~~~\<P^{\mu};L_{a}^{ \lambda_1 ... \lambda_{s}}\rangle  =0,\nn\\
&&~~~~~~\<M^{\mu\nu};L_{a}^{ \lambda_1 ... \lambda_{s}}\rangle =0, \label{poincarecurrent}\\
\nn\\
&&~~~~~~\<L_{a}; L^{\lambda_1}_{b} \rangle  =0,\nn\\
&&~~~~~~\<L^{\lambda_1}_{a}; L^{\lambda_2}_{b} \rangle  = \delta_{ab}~ \eta^{\lambda_1 \lambda_2} ,\nn\\
&&~~~~~~\<L_{a}; L^{\lambda_1\lambda_2}_{b} \rangle  = \delta_{ab}~ \eta^{\lambda_1 \lambda_2} ,\nn\\
&&~~~~~~\<L^{\lambda_1}_{a}; L^{\lambda_2 \lambda_3}_{b} \rangle  =0,\label{currentscalar}\\
&&~~~~~~\<L^{\lambda_1 \lambda_2}_{a}; L^{\lambda_3 \lambda_4}_{b} \rangle  =
\delta_{ab}~ 2!~(\eta^{\lambda_1 \lambda_2}  \eta^{\lambda_3 \lambda_4}+
\eta^{\lambda_1 \lambda_3}  \eta^{\lambda_2 \lambda_4}+
\eta^{\lambda_1 \lambda_4}  \eta^{\lambda_2 \lambda_3}) \nn\\
&&~~~~~~~~~~.....................\nn\\
&&~~~~~~\<L^{\lambda_1...\lambda_n}_{a}; L^{\lambda_{n+1}....\lambda_{2s+1}}_{b} \rangle  = 0,~~~s=0,1,2,3,...\nn\\
&&~~~~~~\<L^{\lambda_1...\lambda_n}_{a}; L^{\lambda_{n+1}....\lambda_{2s}}_{b} \rangle  =
\delta_{ab}~ s!~(\eta^{\lambda_1 \lambda_2}  \eta^{\lambda_3 \lambda_4}...
\eta^{\lambda_{2s-1} \lambda_{2s}} + \textrm{perm}),\nn
\eeqa
The most important conclusion which can be drawn upon above computation is that
Poincar\'e generators $P^{\mu}, M^{\mu\nu}$ are orthogonal to the  gauge  generators
$L_{a}^{ \lambda_1 ... \lambda_{s}}$.

Our intention is to derive the expression for the scalar products, this time
using the invariance of
the scalar product under the action of group transformations which is expressed by the
equation (\ref{symmetic}). New derivation will provide us with an independent
cross check of the direct calculation of the traces of infinite dimensional matrices
which we performed in this section.

\section{\it Group Invariance of the Scalar Product}

Let us first consider the Poincar\'e algebra $L_{\CP}$. The invariant equation for the scalar
product has the form (\ref{symmetic})
\beqa\label{symmetic1}
\<X;[Y, Z]\rangle  + \<[X, Z];Y\rangle =0 ~~~~~~X,Y,Z \in L_G~,\nn
\eeqa
and we shall take
$X=P^{\rho}, Y=M^{\mu\nu}$ and $Z=P^{\lambda}$. The above equation reduces to
$$
\<P^{\rho}; [M^{\mu\nu},P^{\lambda}]\rangle  +\<[P^{\rho},P^{\lambda}];M^{\mu\nu}\rangle =0~,
$$
or using the definition of the commutation relations of the $L_{\CP}$ algebra (\ref{extensionofpoincarealgebra})
we shall get
\be
\eta^{\nu\lambda} \<P^{\rho};P^{\mu}\rangle  - \eta^{\mu\lambda}\<P^{\rho}; P^{\nu}\rangle  =0.
\ee
At $\nu=\lambda=0, \mu=\rho=1$ we shall get $\<P^1;P^1\rangle =0$, at $\nu=\lambda=1, \mu=\rho=0$
we shall get $\<P^0;P^0\rangle =0$, at $\nu=\lambda=2, \mu=1, \rho=0$ we shall get $\<P^0;P^1\rangle =0$ and so on.
This derivation confirms our previous calculation  of the product of momentum generators (\ref{killingform0}).

Taking
$X=P^{\sigma}, Y=M^{\mu\nu}$ and $Z=M^{\lambda\rho}$   the equation (\ref{symmetic}) takes the
form
$$
\<P^{\sigma}; [M^{\mu\nu},M^{\lambda\rho}]\rangle  +\<[P^{\sigma},M^{\lambda\rho}];M^{\mu\nu}\rangle =0~,
$$
or using the $L_{\CP}$ algebra (\ref{extensionofpoincarealgebra})
\beqa
\eta^{\mu\rho} \<P^{\sigma};M^{\nu\lambda}\rangle  - \eta^{\mu\lambda}\<P^{\sigma}; M^{\nu\rho}\rangle +\nn\\
\eta^{\nu\lambda} \<P^{\sigma};M^{\mu\rho}\rangle  - \eta^{\nu\rho}\<P^{\sigma}; M^{\mu\lambda}\rangle + \\
\eta^{\sigma\lambda} \<P^{\rho};M^{\mu\nu}\rangle  - \eta^{\sigma\rho}\<P^{\lambda}; M^{\mu\nu}\rangle =0.\nn
\eeqa
At $\mu=\rho=0, \sigma=1,\nu=2,\lambda=2$ we shall get $\<P^1;M^{23}\rangle =0$, at $\mu=\rho=1, \sigma=0,\nu=2,\lambda=3$
we shall  get $\<P^0;M^{23}\rangle =0$  and so on. This confirms the second equation in (\ref{killingform0}).

Taking
$X=M^{\sigma\delta}, Y=M^{\mu\nu}$ and $Z=M^{\lambda\rho}$, the equation (\ref{symmetic})  takes the
form
$$
\<M^{\sigma\delta}; [M^{\mu\nu},M^{\lambda\rho}]\rangle  +\<[M^{\sigma\delta},M^{\lambda\rho}];M^{\mu\nu}\rangle =0
$$
or using the commutators of the $L_{\CP}$ algebra (\ref{extensionofpoincarealgebra}) we get
\beqa
\eta^{\mu\rho} \<M^{\sigma\delta};M^{\nu\lambda}\rangle  - \eta^{\mu\lambda}\<M^{\sigma\delta}; M^{\nu\rho}\rangle +\nn\\
\eta^{\nu\lambda} \<M^{\sigma\delta};M^{\mu\rho}\rangle  - \eta^{\nu\rho}\<M^{\sigma\delta}; M^{\mu\lambda}\rangle + \\
\eta^{\sigma\rho} \<M^{\delta\lambda};M^{\mu\nu}\rangle  - \eta^{\sigma\lambda}\<M^{\delta\rho}; M^{\mu\nu}\rangle +\nn\\
\eta^{\delta\lambda} \<M^{\sigma\rho};M^{\mu\nu}\rangle  - \eta^{\delta\rho}\<M^{\sigma\lambda}; M^{\mu\nu}\rangle =0.\nn
\eeqa
At $\mu=\rho=0, \sigma=1,\delta=\nu=2,\lambda=3$ we have $\<M^{12};M^{23}\rangle =0$
and so on. This confirms the third  equation in (\ref{killingform0}). Thus all relations
for the scalar product in (\ref{poincare}) are consistent with the requirement of group invariance.

Now let us consider the full algebra $L_G(\CP)$ (\ref{extensionofpoincarealgebra}) .
In the case when $X=M^{\mu\nu}$ and the other two
operators  are spacetime scalars $Y=L_{a}$ and $Y= L_{b}$,
we have to consider the equation
\beqa
 \<M^{\mu\nu};[L_{a},~ L_{b}] \rangle +\<[M^{\mu\nu},~L_{b}];  L_{a} \rangle =0\nn
\eeqa
or using the commutations relations of the algebra (\ref{extensionofpoincarealgebra}) and in
particular that $[M^{\mu\nu},~L_{b}]=0$  the equation takes the form
$$
i f_{abc} \<M^{\mu\nu}; L_c\rangle =0~,
$$
and we can conclude that the product of the operator
$M^{\mu\nu}$ with the spacetime scalar operators $L_c$ is equal to zero
\be\label{first}
\<M^{\mu\nu}; L_c\rangle =0.
\ee
Next we shall consider the product of the $M^{\mu\nu}$ with the spacetime vector operator
$Y= L^{\lambda }_{b}$. For that let consider the case when $X=M^{\mu\nu}$,
$Y=L^{\lambda }_{a}$ and $Y= L_{b}$, so that  we have to consider the equation
$$
 \<M^{\mu\nu};[L^{\lambda}_{a}~ L_{b}] \rangle  +\<[M^{\mu\nu}~L_{b}];  L^{\lambda }_{a} \rangle =0,\nn
$$
or using the definition of the commutators of the algebra (\ref{extensionofpoincarealgebra}) we get
$$
 i f_{abc} \<M^{\mu\nu}; L^{\lambda}_c\rangle    =0,
$$
thus we conclude that the scalar product of the operator
$M^{\mu\nu}$ with the vector operator $L^{\lambda}_c$ is also equal to zero
\be
\<M^{\mu\nu}; L^{\lambda }_c\rangle =0.
\ee
To proceed we have to consider the product of $M^{\mu\nu}$ with the higher rank gauge operator
$L^{\lambda_1...\lambda_s }_{a}$.
Thus let us consider the case when $X=M^{\mu\nu}$, $Y=L^{\lambda_1...\lambda_s }_{a}$ and $Y= L_{b}$
 the invariance  equation is
$$
\<M^{\mu\nu};[L^{\lambda_1...\lambda_s }_{a}~ L_{b}] \rangle
+\<[M^{\mu\nu}~L_{b}]; L^{\lambda_1...\lambda_s }_{a}\rangle =0
$$
and using the definition of the commutations relations of the algebra (\ref{extensionofpoincarealgebra}) we get
\beqa
 \<M^{\mu\nu};~L^{\lambda_1...\lambda_s }_{c}  \rangle  =0,
\eeqa
which is consistent with our previous direct calculation (\ref{poincarecurrent}).

We are interested now to find the scalar products between gauge generators.
In the case when $X=M^{\mu\nu}$,
$Y=L_{a}$ and $Y= L^{\lambda }_{b}$  we have the equation
$$
 \<M^{\mu\nu};[L_{a},~ L^{\lambda}_{b}] \rangle  +\<[M^{\mu\nu}~L^{\lambda }_{b}];  L_{a} \rangle =0
$$
or using the definition of the commutators we get
$$
 i f_{abc} \<M^{\mu\nu}; L^{\lambda}_c\rangle    +
\eta^{\lambda \nu} \< L^{\mu }_{b}; L_{a} \rangle  - \eta^{\lambda\mu} \<L^{ \nu}_{b}~  ; L_{a} \rangle
=0
$$
and because $\<M^{\mu\nu}; L^{\lambda }_c\rangle =0$ we conclude that the scalar product of the operator
$L_{a}$ with the vector operator $L^{\lambda}_c$ is equal to zero
\be
\< L_{a} ; L^{\mu }_{b}\rangle =0 ~~.
\ee
Considering the case $X=M^{\mu\nu}$,
$Y=L_{a}$ and $Y= L^{\lambda_1 \lambda_2 }_{b}$  we have the equation
\beqa
 \<M^{\mu\nu};[L_{a},~L^{\lambda_1 \lambda_2 }_{b}] \rangle  +\<[M^{\mu\nu},~L^{\lambda_1 \lambda_2 }_{b}];  L_{a} \rangle =0,\nn
\eeqa
or
\beqa
if_{abc} \<M^{\mu\nu}; L^{\lambda_1 \lambda_2 }_{c}] \rangle  +~~~~~~~~~~~~~~~~~~~~~~~~~~~~~\nn\\
\eta^{\lambda_1\nu} \<L^{\mu \lambda_2}_{b}; L_{a}\rangle  - \eta^{\lambda_1\mu} \<L^{ \nu\lambda_2}_{b}~  ; L_{a}\rangle
+\eta^{\lambda_2\nu} \<L^{\lambda_1\mu }_{b}; L_{a}\rangle  - \eta^{\lambda_2\mu} \<L^{\lambda_1 \nu}_{b}~  ; L_{a}\rangle
=0 \nn
\eeqa
and because $\<M^{\mu\nu}; L^{\lambda_1 \lambda_2 }_{c}\rangle =0$ we shall get
\beqa
\eta^{\lambda_1\nu} \<L^{\mu \lambda_2}_{b}; L_{a}\rangle  - \eta^{\lambda_1\mu} \<L^{ \nu\lambda_2}_{b}~  ; L_{a}\rangle
+\eta^{\lambda_2\nu} \<L^{\lambda_1\mu }_{b}; L_{a}\rangle  - \eta^{\lambda_2\mu} \<L^{\lambda_1 \nu}_{b}~  ; L_{a}\rangle
=0. \nn
\eeqa
It follows then that
\be
\<L^{\lambda_1 \lambda_2}_{a}; L_{b}\rangle  \sim \eta^{\lambda_1 \lambda_2}.
\ee
Considering the case $X=L^{\lambda_1  }_{a}$,
$Y=L^{\lambda_2  }_{b}$ and $Z=  M^{\mu\nu}$  we have the equation
$$
 \<L^{\lambda_1  }_{a} ;[L^{ \lambda_2 }_{b} ,~M^{\mu\nu}] \rangle
 +\<[L^{\lambda_1  }_{a},~M^{\mu\nu} ]; L^{ \lambda_2 }_{b}  \rangle =0,
$$
or
\beqa
\eta^{\lambda_2\mu} \<L^{\lambda_1}_{a}~  ; L^{\nu}_{b}\rangle -
\eta^{\lambda_2\nu} \<L^{\lambda_1 }_{a}; L^{ \mu }_{b}\rangle +
\eta^{\lambda_1\mu} \<L^{ \nu }_{a}~  ; L^{ \lambda_2}_{b}\rangle -
\eta^{\lambda_1\nu} \<L^{\mu }_{a}; L^{ \lambda_2}_{b}\rangle
=0 \nn
\eeqa
and it follows that
\be
\<L^{\lambda_1 }_{a}; L^{ \lambda_2}_{b}\rangle  \sim \eta^{\lambda_1 \lambda_2}.
\ee
Taking in the last two cases instead of $M^{\mu\nu}$ the operator $L_c$ one can see that
$\<L^{\lambda_1 \lambda_2}_{a}; L_{b}\rangle $ and $\<L^{\lambda_1 }_{a}; L^{ \lambda_2}_{b}\rangle $ are the
isotropic tensors proportional to
$\delta_{ab}$, thus reconfirming the result  (\ref{currentgen}) and (\ref{currentgen1}).
In a similar way one can calculate the structure of the scalar products between the higher rank gauge generators
and confirm that the result (\ref{finale}) for the higher rank generators is fully reproduced.

\section{\it Acknowledgements}
The author would like to thank Ignatios Antoniadis, Ludwig Faddeev and Luis Alvarez-Gaume
for  discussions and   CERN Theory Division, where part of this work was completed,
for hospitality. The author would
like  also to thank M.Vasiliev, E.Ivanov  and J.Buchbinder for helpful discussions
which take place in Dubna workshop "Supersymmetries and Quantum Symmetries".
This work was supported in part by the General Secretariat for Research and Technology of
Greece and the European Regional Development Fund (NSRF 2007-13 ACTION,KRIPIS)

\section{\it Appendix A}

The useful properties of the normalized Hermite polynomials
\be
\psi_m(\xi)={\exp{(-\xi^2/2)} \over \sqrt{2^m m! \sqrt{\pi}}}H_m(\xi)~,~~~~
\psi^{'}_m = \sqrt{2 m} ~\psi_{m-1} -\sqrt{2 (m+1)}\psi_{m+1} ~
\ee
are
\beqa
\langle l\vert m \rangle &=&\int^{\infty}_{-\infty}\psi_l(\xi) \psi_m(\xi) d\xi= \delta_{l,m},\nn\\
\<l\vert {d\over d\xi}\vert m\rangle&=&\int^{\infty}_{-\infty}\psi_l(\xi) \psi^{'}_m(\xi) d\xi=
\sqrt{2 m}~\delta_{l, m-1} -\sqrt{2 (m+1)} \delta_{l, m+1},\nn\\
\<l\vert {\xi}\vert m\rangle &=&\int^{\infty}_{-\infty}\psi_l(\xi) \xi \psi_m(\xi) d\xi=
\sqrt{{m\over 2}}~\delta_{l, m-1} +\sqrt{{(m+1)\over 2}} \delta_{l, m+1},\nn\\
\<l\vert {\xi^2}\vert m\rangle &=&\int^{\infty}_{-\infty}\psi_l(\xi) \xi^2 \psi_m(\xi) d\xi\nn\\
&=& \sqrt{{m\over 2}{m-1\over 2}}~\delta_{l, m-2}+
{(2m+1)\over 2}\delta_{l, m}+\sqrt{{(m+1)\over 2}{(m+2)\over 2}} \delta_{l, m+2}\nn
\eeqa

\end{document}